# Sum of rank ratios: an alternative to percentiles for research assessment, from groundbreaking to mainstream research

Alonso Rodríguez-Navarro

*Departamento de Biotecnología-Biología Vegetal, Universidad Politécnica de Madrid, Avenida Puerta de Hierro 2, 28040, Madrid, Spain*

*E-mail address:* alonso.rodriguez@upm.es ORCID 0000-0003-0462-223X

## Abstract

Assessing research that pushes the boundaries of knowledge is challenging because such work is extremely infrequent, accounting for only about 0.01% of all research outputs. Consequently, knowledge about how to evaluate this type of research is far more limited than the well-established methods used to assess more common research outcomes. This study addresses this gap by using a rank-based approach in which each paper is assigned a unique value equal to the ratio between its local and global ranks. The cumulative value of these ratios, starting from the most cited paper, provides the evaluative basis, and the *Rn*-index described here, using 10 rank ratios, appears to be the best option. Although research assessment based on global ranks was originally developed to evaluate the largest contributors to groundbreaking knowledge—namely, the USA and China, which account for most of the most cited papers—the *Rn*-index has broader applications. This study demonstrates that it is also a better option than the number of top 10% or top 1% highly cited papers, which are the most common indicators used to evaluate countries that seldom or never produce cutting-edge research that pushes the boundaries of knowledge. In all cases, the *Rn*-index reflects the highest-quality science produced by each country. Furthermore, the *Rn*-index can be easily calculated without specialized training in bibliometrics and is insignificantly affected by ties in citation counts.

## 1. Introduction.

Since the mid-20th century, scientific and technological research has played a significant role in the economic growth of countries, becoming as important as capital and labor (Godin, 2004). Consequently, “Government policy-makers, corporate research managers, and university administrators need valid and reliable S&T indicators” (Garfield & Welljams-Dorof, 1992, p. 321). However, this demand has not been fully met. For example, some well-known institutions publish country rankings in which Japan's indicators are comparable to those of developing countries (Pendlebury, 2020), which is clearly inaccurate (Rodríguez-Navarro, 2025b; Rodríguez-Navarro & Brito, 2024a).

The lack of accurate indicators is especially problematic when assessing the infrequent research that produces landmark scientific advances and patents, which is estimated to account for only 0.01% of total scientific output (Bornmann et al., 2018; Poege et al., 2019). This low frequency of landmark papers makes it impossible to assess the contribution, or the probability of making a contribution, of individual countries by simply counting papers. For example, in a specific field—e.g., electric batteries, solar cells, or graphene—with 15,000 papers published per year, the number of groundbreaking papers over a four-year period would be approximately six. Such a small number makes it impossible to rank 20 advanced countries using the criterion of groundbreaking contributions based solely on paper counts. Avoiding this problem, most citation-based indicators are calculated using much more frequent categories of papers. For example, the leading institutions responsible for national research assessments—such as the US National Science Board, the European Commission, and the Australian Strategic Policy Institute—base their evaluations on the number of papers within the top 10% or top 1% most cited papers (hereinafter $P_{top\ 10\%}$ and $P_{top\ 1\%}$).

While there is no doubt about the societal benefits of landmark scientific discoveries—"Scientific breakthroughs and disruptive discoveries are the lifeblood of scientific progress. They push the boundaries of our understanding of the world and drive innovation across multiple fields" (Bornmann et al., 2024; p. 2837)—their evaluation is rarely provided. This limitation has negative consequences in research policy because "You can't manage what you can't measure" and "if you can't measure it, you can't improve it." These statements are credited to Peter Drucker, widely regarded as the father of modern management (Drucker, 1954).

In principle, attempts to develop an indicator that reveals similarities or differences in countries' performance in pushing the boundaries of knowledge can be based on citations. Although the correlation between the scientific relevance of a paper, as judged by peer review, and its number of citations is well established (Rodríguez-Navarro & Brito, 2020; Thelwall et al., 2023; Traag & Waltman, 2019), these studies do not address the level of groundbreaking research. For example, the highest peer-review qualification assigned to papers in the UK Research Excellence Framework—world-leading level—corresponds to the top 1% by citations, or to a broader percentile depending on the field (Rodríguez-Navarro & Brito, 2020). However, the reliability of citation-based indicators for evaluating groundbreaking research is supported by studies based on Nobel Prizes (Rodríguez-Navarro, 2016) and on outstanding technological and commercial outcomes (Poege et al., 2019).

### 1.1. Evaluation of landmark scientific discoveries

As already explained, when considering very narrow top percentiles of global papers—e.g., 0.1% or 0.01%—the research evaluation of countries cannot be reliably performed by simply counting papers. Certainly, based on the mathematical properties of percentile indicators, it is possible to calculate the expected frequency of these infrequent papers (Rodríguez-Navarro & Brito, 2018b). However, these calculations are accurate only in ideal research systems, in which top-percentile counts follow a power-law distribution, whereas deviations from this ideal model are common (Rodríguez-Navarro, 2025b). In countries, such deviations occur when one or a few elite groups are atypically successful relative to the bulk of the country's research output. In these cases, the number of very highly cited papers within narrow top percentiles cannot be predicted from the percentile power law calculated using broader top percentiles.

It is worth noting that previous attempts to evaluate the most cited papers as outliers—e.g., (Glänzel, 2013; Prathap, 2014)—are consistent with the ideal cases that can be addressed using double-rank analysis (Rodríguez-Navarro & Brito, 2019). The real issue here is the deviation from the ideal model that occurs when one or several papers from a single author or a small group of authors receive unexpectedly high citation counts relative to the rest of a country's researchers. If these papers are groundbreaking, their success will have occurred independently of the citation counts of the remaining publications.

The study of domestic publications from the USA and South Korea in the field of solar cells, reported in a previous paper (Rodríguez-Navarro, 2025b), illustrates this issue. The US research system is larger than that of South Korea, with 5474 versus 4193 total papers, 799 versus 285 papers in the top 10%, and 113 versus 29 papers in the top 1%, respectively. Furthermore, the US system appears more efficient, as the ratio between $P_{top\ 1\%}$ and $P_{top\ 10\%}$ is 0.14 for the USA, compared with 0.10 for South Korea. This suggests that, in theory, the US research system should produce significantly more very highly cited papers than that of South Korea.

However, this expectation does not hold true. Figure 1, constructed using the above-mentioned results (Rodríguez-Navarro, 2025b), shows that the expected global ranks of the most cited papers, based on an extension of the power law (closed symbols), deviate substantially from the global ranks of the actual papers (open symbols). Considering the ranks of the 10 most cited papers (Figure 1, right panel), a reasonable conclusion is that landmark research pushing the boundaries of knowledge in solar cells is produced at a similar level in both the USA and South Korea.

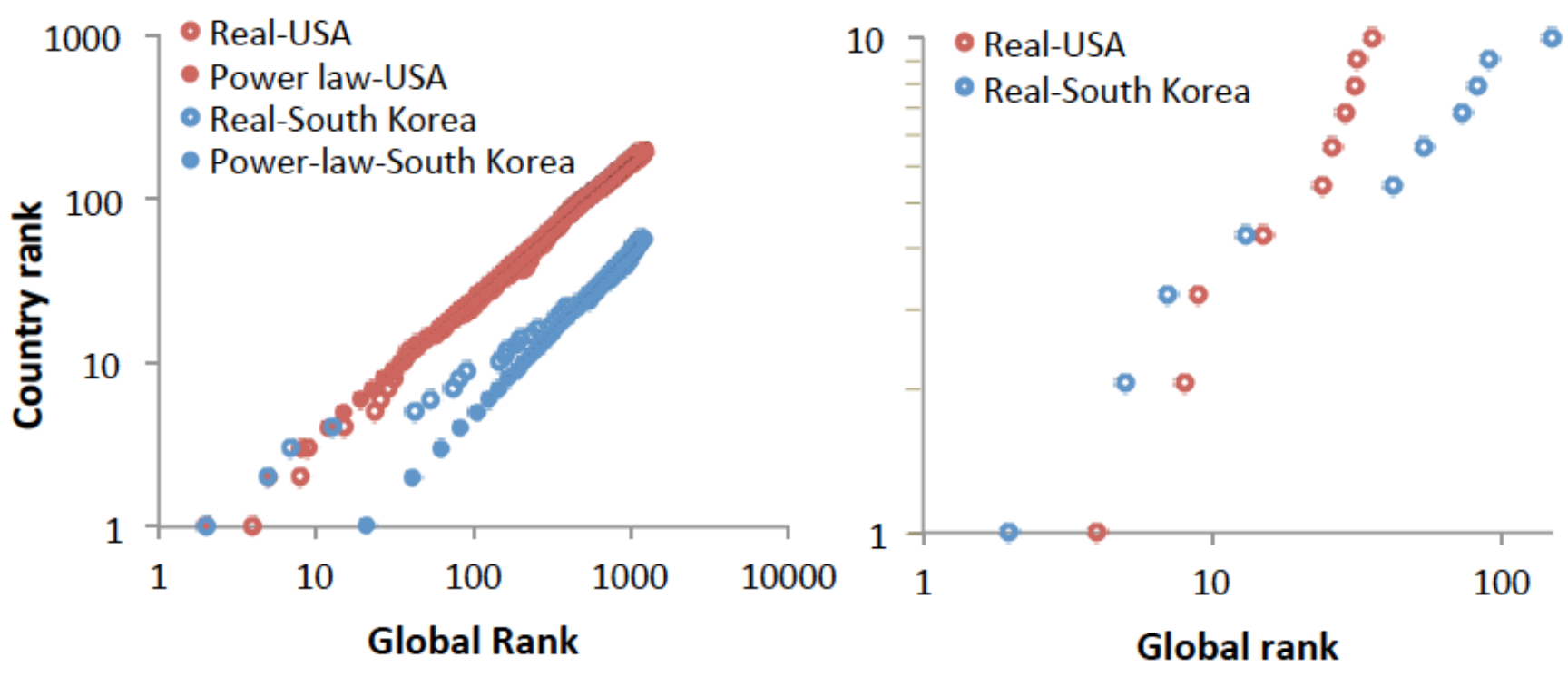

Figure 1. Country versus global rank of domestic papers from the USA and South Korea in solar cells. Comparison of actual global ranks with those calculated using the power law defined by Ptop 10% and Ptop 1%. Left panel: papers within the top 10% by citations. Right panel: the 10 most cited papers from the USA and South Korea. Left panel reproduced from Rodríguez-Navarro (2025b).

## 1.2. The *Rk* and disruption indices

The *Rk*-index was devised specifically to evaluate research that pushes the boundaries of knowledge using synthetic series of lognormally distributed numbers (Rodríguez-Navarro & Brito, 2024b). Like percentile indicators, the *Rk*-index is a nonparametric metric based on the citation-based ranks of papers—ordered from the most cited to the least cited—rather than on their raw citation counts (Conover & Iman, 1981). The key difference between the *Rk*-index and traditional top-percentile indicators is that, in the latter, all papers within a given top percentile are assigned the same value—for instance, all papers in the top 10% are assigned a value of 1, regardless of their individual global ranks. In contrast, the *Rk*-index assigns a distinct value to each paper based on its specific global rank, which is important when the data deviate from the ideal model (Figure 1). The Hazen formula also assigns a distinct value to each paper (Bornmann & Williams, 2020), but it was discarded during the development of the *Rk*-index because it did not yield a convenient indicator (negative unpublished results).

The *Rk*-index is calculated by multiplying by 1000 the geometric mean of the inverses of the global ranks of a country's 10 highest-ranked papers, after adding 20 to each rank (Rodríguez-Navarro & Brito, 2024b). According to this formula, the maximum value of the *Rk*-index is 39.5. One weakness of this index is that, in highly competitive countries engaged in disruptive research, the *Rk*-index can exceed 30, at which point its accuracy diminishes.

A different index that might be used to evaluate research that pushes the boundaries of knowledge is the disruption index (DI) (Funk & Owen-Smith, 2017; Wu et al., 2019), which is based on bibliographic coupling to a focal paper within a given topic. Leibel and Bornmann (2024) discuss the characteristics, variants, and widespread applications of the DI. Despite the interest this index has attracted, it refers exclusively to a subset of the scientific output that pushes the boundaries of knowledge. For example, if the DI were accepted as a measure of landmark research that pushes the boundaries of knowledge, the revolutionary mRNA vaccine-delivery technology and CRISPR technology would be excluded from the category of cutting-edge research, despite their enormous effects on science and society (Kamerlin, 2023).

### 1.3. The combination-addition property

In robust research indicators, if several research systems (with no overlapping publications) are combined, the indicator for the combined system should equal the sum of the indicators of the individual systems. Metrics such as the number of papers or citations, percentile indicators, and number of citations normalized by the mean citation count all fulfill this combination-addition property. However, not all proposed indicators meet this criterion. For example, the *h*-index does not satisfy the combination-addition property due to its mathematical inconsistency (Brito & Rodríguez-Navarro, 2021).

Because the *Rk*-index is equivalent to percentile indicators (Rodríguez-Navarro & Brito, 2024b), it should theoretically satisfy the combination-addition property. However, due to its loss of accuracy in highly competitive countries (as explained above), in practice, it may fail in these countries. For instance, the *Rk*-index values for domestic and collaborative US papers in the field of stem cells are 32.2 and 23.2, respectively (Rodríguez-Navarro, 2025a). A reliable indicator should assign a value close to the sum of these two indicators (i.e., 55.4) when both sets of papers are combined. However, this value exceeds the maximum possible value of the *Rk*-index (39.5), demonstrating the weakness of the metric at high-performance levels.

## 2. Aim of this study

As previously discussed, evaluating cutting-edge research that pushes the boundaries of knowledge is critically important for policymakers, corporate research managers, and university administrators, who need to understand how the research output of their countries or institutions compares with that of global competitors. The *Rk*-index was designed for this purpose (Rodríguez-Navarro & Brito, 2024b). However, the extensive application of the index to many countries (Rodríguez-Navarro, 2025a) revealed weaknesses, especially in the evaluation of the major scientific producers—the USA and China. Furthermore, the *Rk*-index does not satisfy the combination-addition property.

These weaknesses of the *Rk*-index prompted the development of a similar indicator free from these limitations. Subsequently, this study aimed to examine the applicability of the same indicator for evaluating the best science produced by countries that seldom or never publish cutting-edge research that pushes the boundaries of knowledge. In principle, this should be possible because a previous study using synthetic series of lognormally distributed numbers demonstrated the high versatility of the *Rk*-index, showing that it is proportional to very narrow top percentiles at high values and to wider top percentiles at lower values (Rodríguez-Navarro & Brito, 2024b).

## 3. Materials and methods

This study builds upon two previous papers (Rodríguez-Navarro, 2025a; Rodríguez-Navarro & Brito, 2024b). Accordingly, all materials and methods used here are the same as those employed in those prior studies. The reported *Rk*-index values are taken directly from (Rodríguez-Navarro, 2025a), while the new *Rn*-index values described below were calculated from the raw data used in that same paper. A brief description of the methods is as follows:

The 600 lognormally distributed synthetic series used in this study (Brito & Rodríguez-Navarro, 2019, 2021) were generated using a single $\sigma$ value of 1.1 (Radicchi et al., 2008; Rodríguez-Navarro & Brito, 2018a; Thelwall, 2016; Viiu, 2018), with $\mu$ values ranging from 4.0 to 2.0. For each $\mu$ value, three series were created with 800, 400, and 200 numbers, simulating numbers of citations. Global publication output was simulated using a combined dataset of 280,000 numbers derived from the 600 individual series. Rank analyses were performed by ordering these 280,000 numbers from highest to lowest. To assign global ranks to individual series, the ranks of each series' numbers were located within the combined set of 600 series.

Country-level publication data were retrieved from the Science Citation Index Expanded database in Clarivate's Web of Science Core Collection, using the Advanced Search tool. The Citation Report tool was then used to retrieve both publications and their annual citation counts. The publication window was 2014–2017 but the citation window was delayed to 2019–2022 for the reasons given in (Wang et al., 2017). To simplify the search process, global searches were limited to the 75 countries with the highest number of publications and citations (according to Clarivate's InCites), plus Latvia and Malta to complete the 27 EU countries. The UK was not included in the EU. Searches were restricted to "articles" (DT=Articles).

## 4. Results

### 4.1. The *Rn*-index

After testing several alternative formulas to the *Rk*-index, I found that summing the rank ratios of the top-ranked papers—defined as the country rank divided by the global rank—produced a more robust indicator. I defined the *Rn*-index as the result of multiplying the sum of these rank ratios by 10.

Table 1. Paper ranks and calculation of the *Rk* and *Rn* indices for US papers in lithium batteries[a]

| Country rank | Global rank | | | 1/(20+Global Rank) | | | Country rank/Global rank | | |
|---|---|---|---|---|---|---|---|---|---|
| | Dom[b] | Collab[b] | All[b] | Dom[b] | Collab[b] | All[b] | Dom[b] | Colla[b] | All[b] |
| 1 | 2 | 1 | 1 | 0.045 | 0.048 | 0.048 | 0.500 | 1.000 | 1.000 |
| 2 | 3 | 5 | 2 | 0.043 | 0.040 | 0.045 | 0.667 | 0.400 | 1.000 |
| 3 | 7 | 9 | 3 | 0.037 | 0.034 | 0.043 | 0.429 | 0.333 | 1.000 |
| 4 | 8 | 13 | 5 | 0.036 | 0.030 | 0.040 | 0.500 | 0.308 | 0.800 |
| 5 | 10 | 19 | 7 | 0.033 | 0.026 | 0.037 | 0.500 | 0.263 | 0.714 |
| 6 | 14 | 37 | 8 | 0.029 | 0.018 | 0.036 | 0.429 | 0.162 | 0.750 |
| 7 | 15 | 39 | 9 | 0.029 | 0.017 | 0.034 | 0.467 | 0.179 | 0.778 |
| 8 | 17 | 41 | 10 | 0.027 | 0.016 | 0.033 | 0.471 | 0.195 | 0.800 |
| 9 | 20 | 48 | 13 | 0.025 | 0.015 | 0.030 | 0.450 | 0.188 | 0.692 |
| 10 | 21 | 61 | 14 | 0.024 | 0.012 | 0.029 | 0.476 | 0.164 | 0.714 |
| ***Rk*-index[c]** | | | | **32.2** | **23.2** | **37.2** | | | |
| ***Rn*-index[d]** | | | | | | | **48.9** | **31.9** | **82.5** |
| **Geometric mean of rank ratios* 100** | | | | | | | **48,5** | **26.8** | **81.7** |

[a] Data are taken from Rodríguez-Navarro, 2025a.
[b] Domestic (Doc), collaborative (Collab), and all papers (All).
[c] The *Rk*-index is the geometric mean of the data in the column multiplied by 1000.
[d] The *Rn*-index is the sum of the data in the column multiplied by 10.

Table 1 presents the global and country ranks of the 10 most cited US publications—both domestic and collaborative, independently and combined—in the field of lithium batteries along with the corresponding calculations of both the *Rk* and *Rn* indices. Because domestic and collaborative publications do not share common papers, a robust research indicator should satisfy the combination-addition property. In the example shown in Table 1, the *Rn*-index deviates by only 2% from this property, whereas the *Rk*-index shows a much larger deviation of 49%. Table 1 also includes the geometric means of the rank ratios.

Next, I compared the *Rk* and *Rn* indices across a selection of the synthetic series previously used to calculate the *Rk*-index (Rodríguez-Navarro & Brito, 2024b). At low values, the *Rk*-index is approximately double the *Rn*-index, but at higher values the two indices converge (not shown results).

To compare the *Rk* and *Rn* indices in real cases—and at higher values than those tested with the synthetic series—I used data reported in a previous study across five technological fields (Table 2 in Rodríguez-Navarro, 2025a), focusing on cases in which *Rk*-index values were not much lower than 2.0. This threshold was chosen because meaningful differences between the *Rk* and *Rn* indices emerge only at high index values.

Figure 2 (numerical data provided in Supplementary Table S1) shows that, at low values of the *Rk*-index (approximately < 5), the *Rk*-index is roughly double the *Rn*-index. As the values increase, the two indices remain correlated, but the *Rn*-index grows more rapidly than the *Rk*-index. Differences between the two indices typically arise when the first one or two papers have exceptionally low ranks that are substantially separated from the ranks of the subsequent papers. For example, in the case of Japan in semiconductors, Japan published the top-ranked global paper, whereas its next three papers are ranked 95, 99, and 111.

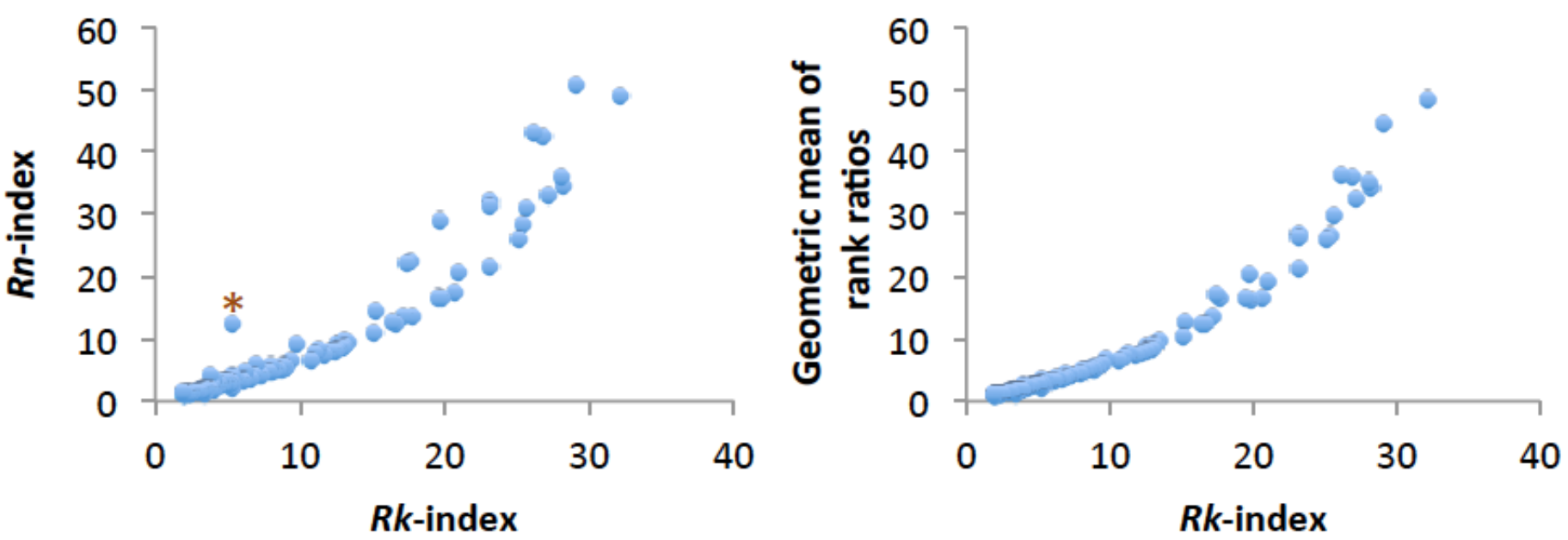

Figure 2. Comparison of the *Rn* and *Rk* indices calculated across 17 countries and five technological topics (Supplementary Table S1). Because the *Rn*-index is the sum of rank rations, the geometric mean of the same rank ratios is included for comparison. Asterisk, Japan in semiconductors.

The faster increase of the *Rn*-index relative to the *Rk*-index improves the evaluation of the major scientific producers—the USA and China—because of the limited capacity of the *Rk*-index to accurately assess these cases (Table 1). Consequently, the *Rn*-index satisfies the combination-addition property better than the *Rk*-index in the major scientific producers. Table 2 presents the cases of the USA, China, the EU, and two developed countries: Australia and South Korea. As a general rule, at low values of the *Rn* and *Rk* indices, deviations from this property are only slightly smaller for the *Rn*-index than for the *Rk*-index, with a maximum deviation of 25%. In contrast, for the major scientific producers—the USA and China—deviations from the combination-addition property are minimal for the *Rn*-index but very high for the *Rk*-index, reaching around 50%.

Table 2. Values of the *Rk* and *Rn* indices in major research systems—the USA, China, and the EU—and in two additional countries, Australia and South Korea, as examples of advances countries. For each country and technology, deviation from the combination-addition property is shown by the ratio of the sum of the domestic and collaboration indices to the index for all papers[a]

| Country | Graphene | | Semiconductors | | Solar cells | | Lithium batteries | | Composite materials | |
|---|---|---|---|---|---|---|---|---|---|---|
| | ***Rk*** | ***Rn*** | ***Rk*** | ***Rn*** | ***Rk*** | ***Rn*** | ***Rk*** | ***Rn*** | ***Rk*** | ***Rn*** |
| USA (D) | 25.7 | 30.9 | 28.3 | 34.5 | 25.1 | 25.9 | 32.2 | 48.9 | 28.1 | 33.3 |
| USA (IC) | 29.1 | 50.8 | 25.4 | 28.1 | 19.8 | 16.4 | 23.2 | 31.9 | 27.2 | 32.9 |
| USA (all) | 37.6 | 86.9 | 33.4 | 52.9 | 29.2 | 36.1 | 37.2 | 82.5 | 36.2 | 68 |
| USA (D)+(IC)/(all) | **1.46** | **0.94** | **1.61** | **1.18** | **1.54** | **1.17** | **1.49** | **0.98** | **1.53** | **0.97** |
| China (D) | 23.2 | 21.4 | 15.3 | 14.4 | 13.1 | 9.8 | 20.7 | 17.5 | 19.6 | 16.5 |
| China (IC) | 26.9 | 42.5 | 21.0 | 20.6 | 15.2 | 10.8 | 23.2 | 31.2 | 26.2 | 43.1 |
| China (all) | 32.0 | 55.5 | 27.2 | 34.8 | 17.4 | 20.6 | 27.6 | 38.6 | 32.6 | 59.6 |
| China (D)+(IC)/(all) | **1.57** | **1.15** | **1.33** | **1.01** | **1.63** | **1.00** | **1.59** | **1.26** | **1.40** | **1.00** |
| EU (D) | 11.1 | 7.5 | 13.1 | 8.8 | 7.3 | 4.1 | 9.5 | 6.3 | 13.1 | 8.9 |
| EU (IC) | 13.0 | 8.4 | 17.2 | 13.5 | 17.2 | 12.3 | 12.6 | 9.0 | 17.8 | 13.5 |
| EU (all) | 18.9 | 14.9 | 21.6 | 19.3 | 17.5 | 13.3 | 17.8 | 14.8 | 22.3 | 19.6 |
| EU (D)+(IC)/(all) | **1.28** | **1.07** | **1.40** | **1.16** | **1.40** | **1.23** | **1.24** | **1.03** | **1.39** | **1.14** |
| Australia (D) | 3.8 | 1.9 | 1.2 | 0.6 | 2.2 | 1.1 | 2.2 | 1.1 | 1.8 | 0.9 |
| Australia (IC) | 6.6 | 3.6 | 4.5 | 2.3 | 3.6 | 1.9 | 4.5 | 2.4 | 5.0 | 2.6 |
| Australia (all) | 8.3 | 4.7 | 5.3 | 2.8 | 4.7 | 2.5 | 5.2 | 2.8 | 5.4 | 2.9 |
| Australia (D)+(IC)/(all) | **1.26** | **1.17** | **1.07** | **1.04** | **1.23** | **1.19** | **1.28** | **1.25** | **1.27** | **1.23** |
| South Korea (D) | 3.6 | 1.8 | 3.8 | 1.9 | 17.4 | 22.1 | 4.8 | 2.5 | 3.9 | 2.2 |
| South Korea (IC) | 7.0 | 5.9 | 12.2 | 7.9 | 9.0 | 5.2 | 4.2 | 2.1 | 5.3 | 2.8 |
| South Korea (all) | 9.1 | 7.0 | 13.5 | 9.0 | 21.2 | 25.6 | 7.0 | 3.8 | 7.7 | 4.5 |
| S Korea (D)+(IC)/(all) | **1.16** | **1.10** | **1.18** | **1.09** | **1.24** | **1.07** | **1.28** | **1.20** | **1.20** | **1.12** |

[a] Values of the *Rk*-index for domestic and international collaborations are taken from Rodríguez-Navarro, 2025a.

[b] Abbreviations: (D) domestic papers; (IC) international collaborative papers; (all) all papers.

Another advantage of the *Rn*-index is its low sensitivity to papers tied in citation counts, which can introduce uncertainty in top-percentile evaluations (Schreiber, 2013). In large, competitive countries, tied publications are unlikely, even for the paper ranked tenth. In less competitive countries, tied publications can occur, but they have an insignificant effect on the *Rn*-index. For example, among domestic publications from Finland in graphene research in 2017 (see below, Section 4.3), the publication with a country rank of five has 40 citations; among world publications, there are four papers ranked 3757–3760 with 40 citations. Consequently, the maximum difference between the rank ratios, depending on the global rank considered, is 0.08%.

### 4.2. Variants of the *Rn*-index

Two types of variants might improve the functionality of the *Rn*-index. The first concerns the number of publications used in the formula of the index—10 in the calculations presented so far—and the second aims to reduce the weight of the rank ratios of the top-ranked papers, which are 1 and 0.5 for the first- and second-ranked papers, respectively, in the corresponding countries. For example, such a modification would reduce the value of the *Rn*-index shown for Japan in the left panel of Figure 2.

Regarding the number of papers included in the formula, using 10 papers to calculate the *Rn*-index is only one option. To investigate the possibility of using a different number of papers, I calculated the cumulative values of the rank ratios for the top 30 papers across four technologies, including both large research systems (USA, China, and the EU) and smaller producers (Figure 3). In many cases, the plots are nearly linear—for example, for domestic papers from China in graphene, the EU in semiconductors, and the UK in solar cells. However, in other cases, the cumulative rank ratios of the most cited papers deviate from the linear trend fitted to the remaining data—for example, for domestic papers from the USA in graphene, South Korea in solar cells, and France in composite materials.

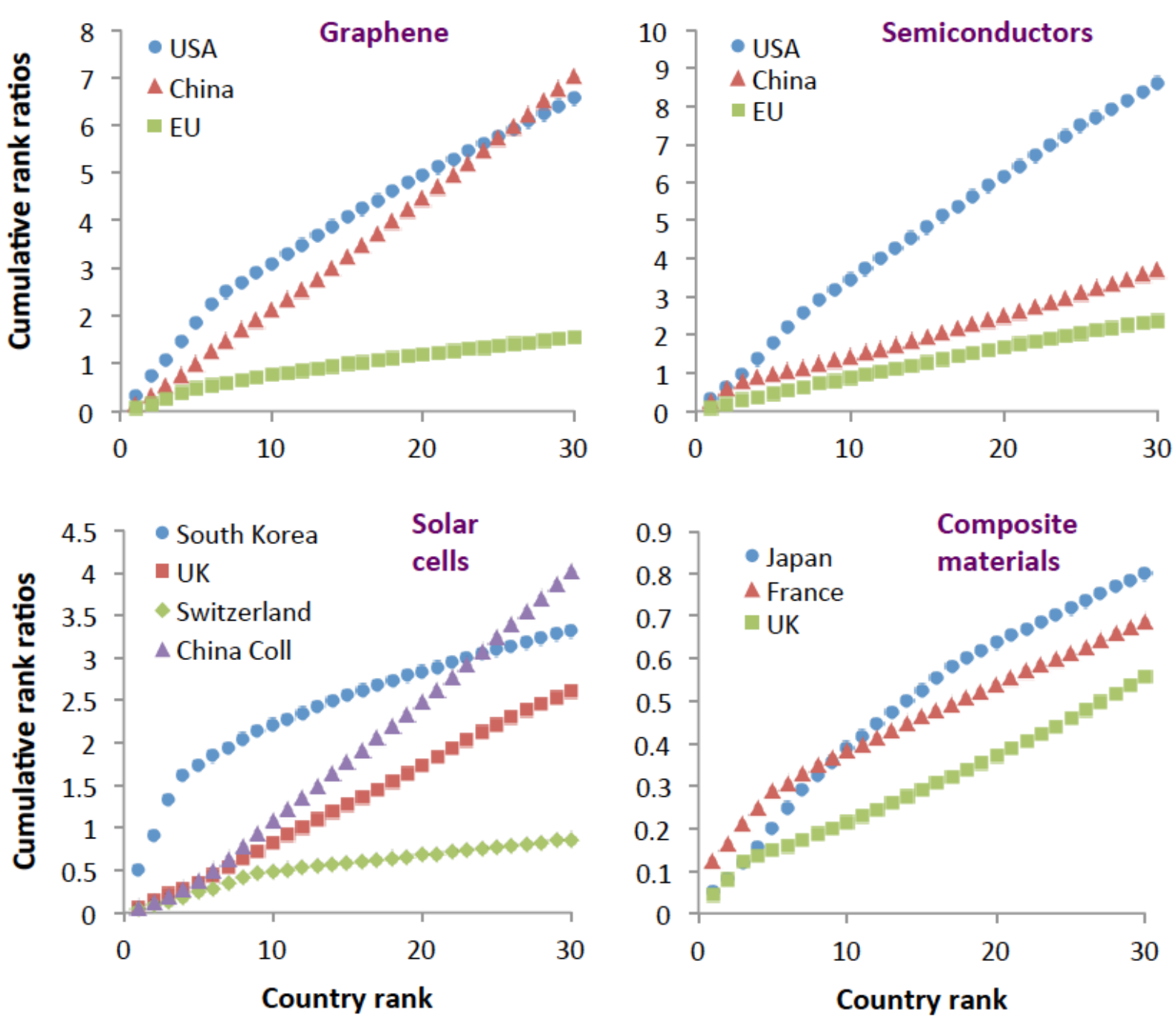

Figure 3. Cumulative values of rank ratios versus country rank of papers in representative cases of countries and technological fields. The *Rn*-index corresponds to the cumulative rank ratios up to the paper ranked 10th.

Based on the analysis of numerous cases—such as those shown in Figure 3—it can be concluded that using more than 10 papers would decrease the capacity of the index to evaluate the best science of a country. Conversely, using only five papers may provide a reliable alternative to the Rn-index based on 10 papers.

To test this alternative, I compared the *Rn*-indices based on 10 and five papers across 17 countries and five technologies (Table 3). For better comparability, the *Rn*-index based on five papers was calculated by summing the rank ratios and multiplying the result by 20 to match the scale of the 10-paper variant.

The first conclusion drawn from the data in Table 3 is that the proportion of cases where the *Rn*-indices calculated from five and 10 papers differ substantially is low; in only 47 out of 170 cases does the difference exceed 25%. If two *Rn*-indices are nearly identical, it implies that the plot of cumulative rank ratios is a straight line—as seen with domestic papers from China in graphene, the EU in semiconductors, and the UK in solar cells (Figure 3).

In contrast, significant differences between the five- and 10-paper *Rn*-indices—those exceeding 25%—indicate that the first data points of the cumulative plot (Figure 3) notably deviate from a straight line. In most of these cases (41 out of the 47), the *Rn*- index for five papers is higher than that for 10 papers. A representative example is domestic papers from South Korea in solar cells. In the remaining six cases, the five-paper *Rn*-index is lower than the 10-paper version; a model example of these cases are collaborative papers from China in solar cells.

Table 3. Values of two variants of the *Rn*-index: calculated from the global ranks of the 10 and five most cited papers, across several countries and five research fields.

| Country[a] | Graphene | | Semiconductors | | Solar cells | | Li batteries | | Composite materials | |
|---|---|---|---|---|---|---|---|---|---|---|
| | 10 | 5 | 10 | 5 | 10 | 5 | 10 | 5 | 10 | 5 |
| USA (D) | 30.9 | 37.0 | 34.5 | 36.1 | 25.9 | 26.2 | 48.9 | 51.9 | 35.8 | 40.4 |
| USA (IC) | 50.8 | 70.9 | 28.1 | 25.2 | 16.4 | 15.3 | 31.9 | 46.1 | 32.9 | 38.3 |
| China (D) | 21.4 | 20.0 | 14.4 | 19.4 | 9.8 | 12.3 | 17.5 | 12.2 | 16.5 | 16.4 |
| China (IC)) | 42.5 | 56.7 | 20.6 | 22.0 | 10.8 | 7.7 | 31.2 | 42.7 | 43.1 | 63.1 |
| EU (D) | 7.5 | 9.6 | 8.8 | 9.1 | 4.1 | 3.4 | 6.3 | 7.0 | 8.9 | 9.3 |
| EU (IC)) | 8.4 | 7.4 | 13.5 | 13.8 | 12.3 | 11.5 | 9.0 | 10.2 | 13.5 | 12.1 |
| South Korea (D) | 1.8 | 1.8 | 1.9 | 1.9 | 22.1 | 34.7 | 2.5 | 2.4 | 2.2 | 2.4 |
| South Korea (IC) | 5.9 | 8.8 | 7.9 | 8.2 | 5.2 | 4.6 | 2.1 | 2.0 | 2.8 | 3.2 |
| UK (D) | 0.8 | 0.9 | 4.6 | 4.1 | 8.2 | 7.1 | 1.0 | 1.0 | 2.1 | 2.9 |
| UK (IC) | 5.2 | 4.9 | 12.6 | 12.3 | 22.3 | 32.2 | 1.6 | 1.4 | 5.1 | 4.8 |
| Japan (D) | 2.3 | 4.0 | 12.3 | 22.2 | 2.3 | 2.4 | 2.6 | 3.4 | 3.9 | 4.0 |
| Japan (IC) | 3.4 | 2.7 | 6.5 | 6.1 | 9.0 | 12.8 | 3.5 | 5.2 | 3.9 | 3.4 |
| Singapore (D) | 4.2 | 4.9 | 0.9 | 0.7 | 1.0 | 0.7 | 1.5 | 1.4 | 0.9 | 0.8 |
| Singapore (IC) | 4.5 | 3.8 | 3.9 | 5.4 | 2.6 | 2.7 | 5.0 | 5.4 | 5.5 | 6.4 |
| Germany (D) | 0.5 | 0.6 | 1.7 | 1.8 | 1.2 | 1.0 | 3.1 | 2.2 | 1.4 | 1.4 |
| Germany (IC) | 8.1 | 7.9 | 8.2 | 9.8 | 5.6 | 6.4 | 9.3 | 11.5 | 7.4 | 6.6 |
| Switzerland (D) | 0.2 | 0.3 | 1.4 | 1.9 | 4.7 | 4.6 | 0.3 | 0.3 | 1.2 | 2.0 |
| Switzerland (IC) | 1.2 | 1.5 | 5.3 | 6.5 | 28.9 | 43.4 | 0.4 | 0.4 | 2.2 | 2.6 |
| Canada (D) | 1.0 | 0.6 | 0.6 | 0.6 | 1.4 | 2.1 | 4.2 | 7.1 | 1.3 | 1.6 |
| Canada (IC) | 2.6 | 3.6 | 3.1 | 3.2 | 5.6 | 6.8 | 4.6 | 6.9 | 2.8 | 3.6 |
| Australia (D) | 1.9 | 1.9 | 0.6 | 0.7 | 1.1 | 1.2 | 1.1 | 1.0 | 0.9 | 0.8 |
| Australia (IC) | 3.6 | 3.0 | 2.3 | 2.3 | 1.9 | 1.4 | 2.4 | 1.8 | 2.6 | 2.4 |
| France (D) | 0.8 | 1.3 | 0.8 | 0.8 | 0.4 | 0.3 | 0.9 | 0.9 | 3.8 | 5.8 |
| France (IC) | 1.9 | 2.6 | 3.1 | 2.6 | 2.7 | 3.4 | 2.0 | 1.9 | 1.3 | 1.2 |
| India (D) | 0.4 | 0.4 | 1.0 | 0.8 | 0.8 | 0.6 | 0.4 | 0.3 | 1.8 | 1.4 |
| India (IC) | 0.7 | 0.5 | 1.0 | 0.8 | 0.9 | 1.1 | 0.5 | 0.6 | 3.1 | 4.1 |
| Italy (D) | 0.2 | 0.2 | 1.3 | 2.1 | 1.1 | 1.5 | 0.2 | 0.2 | 1.0 | 1.1 |
| Italy (IC) | 1.7 | 2.1 | 2.8 | 2.9 | 4.1 | 3.9 | 1.2 | 1.9 | 2.3 | 2.3 |
| Spain (D) | 0.5 | 0.5 | 0.9 | 1.3 | 0.7 | 0.6 | 0.2 | 0.2 | 1.2 | 1.3 |
| Spain (IC) | 3.2 | 4.8 | 3.0 | 4.0 | 2.2 | 1.8 | 1.2 | 1.4 | 1.5 | 1.2 |
| Netherlands (D) | 0.4 | 0.5 | 0.6 | 0.7 | 0.3 | 0.3 | 0.2 | 0.3 | 0.4 | 0.5 |
| Netherlands (IC) | 2.9 | 4.5 | 3.0 | 4.3 | 1.9 | 1.8 | 0.4 | 0.2 | 1.2 | 1.3 |
| Sweden (D) | 0.1 | 0.1 | 0.2 | 0.1 | 0.4 | 0.4 | 0.5 | 0.6 | 0.3 | 0.5 |
| Sweden (IC) | 1.1 | 1.3 | 2.0 | 2.7 | 2.5 | 2.8 | 0.8 | 0.8 | 0.7 | 0.7 |

[a] Abbreviations: (D), domestic; (IC), international collaborations. Five-version, sum of ratios multiplied by 20.

Straight lines in cumulative plots of rank ratios imply homogeneity in the research system, whereas deviations from linearity indicate a mixed system. For example, they may reflect the existence of a small number of elite groups that stand above the general research system, as in the case of South Korea in Figures 1 and 3. Based on this, one might expect significant differences between the five and 10 versions to occur more frequently in collaborative publications than in domestic ones. This expectation is partially supported: among the 41 cases in which the five-paper *Rn*-index is at least 25% higher than the 10-paper version, 26 involve collaborative papers. However, considering the full set of 170 cases (Table 3), the overall proportion remains low. This suggests that many international collaborations produce homogeneous research systems.

The high values observed in the five-paper variant of the *Rn*-index in some cases raise the question of whether this variant still satisfies the combination-addition property. For instance, the domestic and collaborative *Rn*-index values for graphene research in the USA are 37 and 71, respectively, and for lithium batteries they are 52 and 46. The combined *Rn*-indices in these cases are 100—five US papers—and 90, compared with sums of 108 and 98, respectively. These differences—approximately 10% in these two extreme cases—do not appear to represent significant deviations from the combination-addition property.

Table 4. Comparison of the *Rn*-index and the geometric mean of rank ratios multiplied by 1000 in eight representative cases

| **Country**[a] | **Field** | ***Rn*-index** | **Geo. mean** |
|---|---|---|---|
| USA (IC) | Graphene | 50.8 | 44.5 |
| China (IC) | Graphene | 42.5 | 35.9 |
| China (D) | Semiconductors | 14.4 | 12.6 |
| Japan (D) | Semiconductors | 12.2 | 3.6 |
| South Korea (D) | Solar cells | 22.1 | 17.1 |
| UK (IC) | Solar cells | 22.3 | 16.4 |
| Switzerland (IC) | Solar cells | 28.9 | 20.1 |
| China (IC) | Composite materials | 43.1 | 36.2 |

[a] (D) domestic papers, (IC) international collaborative papers

Regarding the reduction in the weight that a paper with a very low rank has in the value of the *Rn*-index, as in the case of Japan in semiconductors discussed above (Figure 2), the use of the geometric mean would produce such a reduction. However, this reduction would occur in all countries (Table 4 shows a few examples), which raises the question of whether it is reasonable, as discussed below.

### 4.3. The *Rn*-index versus top percentile indicators

It is possibly universally accepted that, in citation-based research evaluations, the number of papers in narrow top percentiles—10% or 1%—reveals the success of scientific research (e.g., Bornmann & Marx, 2013); for the European Commission, these numbers reveal scientific excellence: "Scientific excellence is measured by the share of the top 1% and top 10% of the most cited publications" (European Commission, 2024, p. 151). However, this assumption does not hold true because the results of comparisons between countries can be contradictory depending on the selected percentile (Rodríguez-Navarro & Brito, 2019).

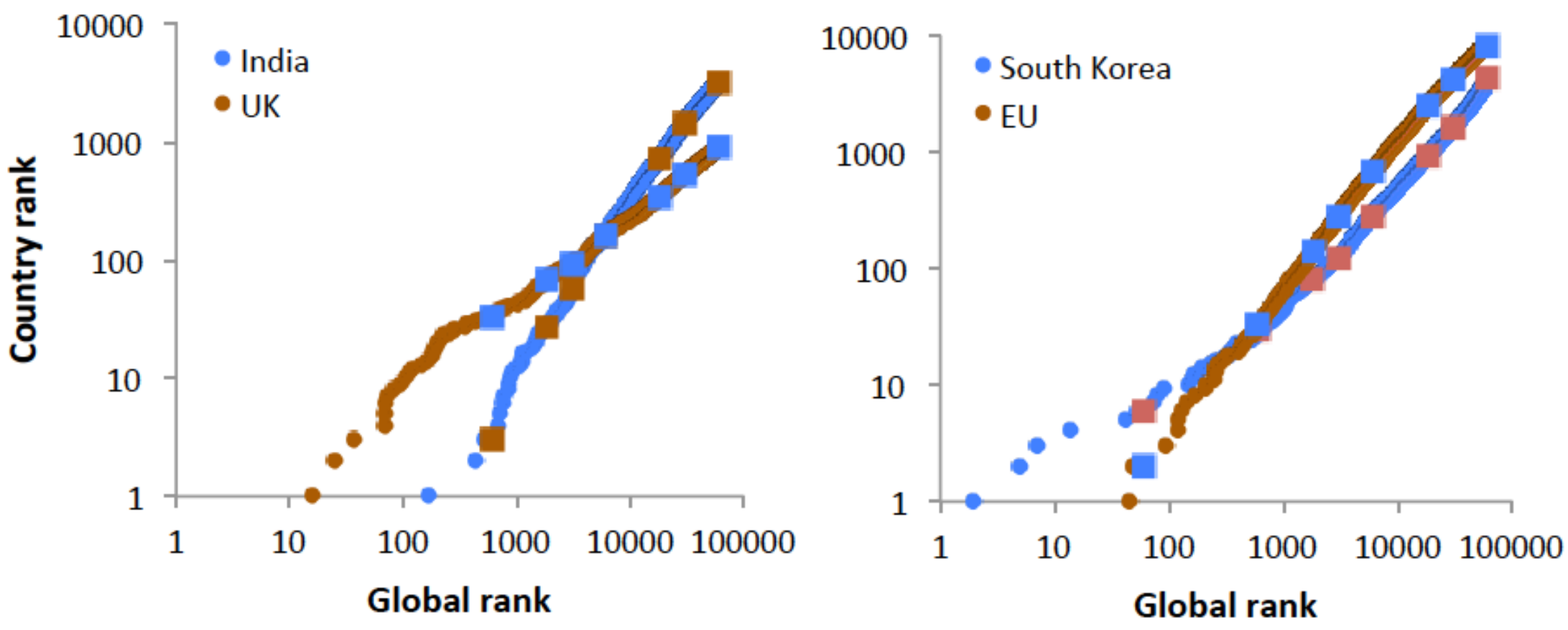

Figure 4. Country versus global rank of domestic papers from India and the UK (left panel), and the USA and South Korea (right panel), in the field of solar cells. Square symbols indicate percentile thresholds: 100, 50, 30, 10, 5, 3, and 1 in the left panel; the same percentiles plus 0.1 in the right panel.

To demonstrate once more the contradictions that appear with the use of percentile-based research evaluations, Figure 4 shows two comparisons of countries in the field of solar cells using double-rank plots, in which top percentiles are shown. The left panel depicts a comparison between domestic publications from India and the UK. India publishes many more papers than the UK—3,138 versus 892—and both countries would be evaluated similarly according to $P_{top\ 10\%}$—163 versus 162—but if the evaluation were based on $P_{top\ 1\%}$, the UK would be evaluated much more favorably—33 versus 3.

The right panel of Figure 4 shows the comparison of domestic publications from South Korea and the EU in the same field. For all top-percentile indicators up to the top 1%, the EU outperforms South Korea, although with large differences between percentiles—671 versus 278 at the top 10% and 32 versus 29 at the top 1%. In contrast, at the top 0.1%, South Korea outperforms the EU—6 versus 2.

Although the data presented in Figure 4 demonstrate inconsistencies in the use of percentiles in research evaluations, $P_{top\ 1\%}$ is the most commonly used percentile indicator in the evaluation of countries, under the assumption that it reveals research success (e.g., King, 2004; National Institute of Science and Technology, 2022; Wagner et al., 2022). This extensive use raises questions about the differences between evaluations based on $P_{top\ 1\%}$ and the *Rn*-index. To address this issue, I calculated $P_{top\ 1\%}$ and the *Rn*-index for 80 selected cases from a previous publication (Rodríguez-Navarro, 2025a). Figure 5 (numerical data in Supplementary Table S2) depicts the scatter plot of these values and shows that the two indicators are not equivalent. Although the values of the two indicators are correlated (Panel A, log-log plot), the dispersion of the data, even at low indicator values (Panel C), is too high to conclude that the two indicators reveal the same characteristic across all research systems and at every level of research success.

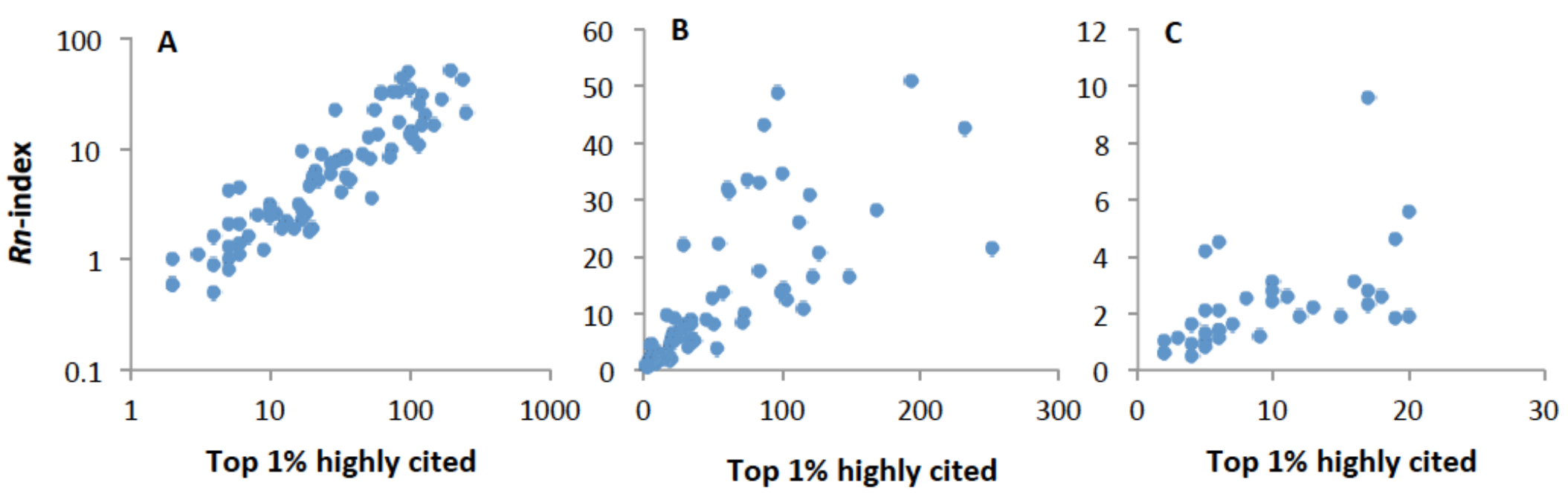

Figure 5. *Rn*-index versus the number of top 1% highly cited papers in 80 selected cases of countries and research fields (Supplementary Table S2). Panel A shows a log-log plot.

Although the top 10% may seem too broad a percentile to assess research excellence, in some cases $P_{top\ 10\%}$ is used to evaluate breakthrough research when $P_{top\ 1\%}$ values are too low (Gaida et al., 2023; Leung et al., 2024). Therefore, to further investigate the differences among $P_{top\ 1\%}$, $P_{top\ 10\%}$, and the *Rn*-index, I compared the values of these three indicators for the two countries that dominate technological sciences—the USA and China (Table 5).

Table 5. Comparison of the *Rn*-index with the numbers of top 10% and top 1% highly cited papers from the USA and China for domestic (D) and international collaborative (IC) papers. The last two columns show the ratios between the compared values

| **Indicator** | **USA (D)** | **USA (IC)** | **China (D)** | **China (IC)** | **USA (D)/ China (D)** | **USA (IC)/ China (IC)** |
|---|---|---|---|---|---|---|
| Graphene | | | | | | |
| **N (all papers)** | 4677 | 7303 | 32745 | 8868 | 0.14 | 0.82 |
| **top 10%** | 603 | 1258 | 3350 | 1676 | 0.18 | 0.75 |
| **top 1%** | 120 | 193 | 252 | 232 | 0.48 | 0.83 |
| ***Rn*-index** | 30.9 | 50.8 | 21.4 | 42.5 | 1.44 | 1.20 |
| Semiconductors | | | | | | |
| **N (all papers)** | 6019 | 5483 | 12183 | 4312 | 0.49 | 1.27 |
| **top 10%** | 832 | 916 | 1352 | 866 | 0.62 | 1.06 |
| **top 1%** | 100 | 169 | 102 | 127 | 0.98 | 1.33 |
| ***Rn*-index** | 34.5 | 28.1 | 14.4 | 20.6 | 2.40 | 1.36 |
| Solar cells | | | | | | |
| **N (all papers)** | 5505 | 5108 | 12883 | 4571 | 0.43 | 1.12 |
| **top 10%** | 773 | 873 | 1116 | 861 | 0.69 | 1.01 |
| **top 1%** | 113 | 149 | 73 | 115 | 1.55 | 1.30 |
| ***Rn*-index** | 25.9 | 16.4 | 9.8 | 10.8 | 2.64 | 1.52 |
| Lithium batteries | | | | | | |
| **N (all papers)** | 3376 | 2921 | 12577 | 3814 | 0.27 | 0.77 |
| **top 10%** | 584 | 521 | 983 | 687 | 0.59 | 0.76 |
| **top 1%** | 97 | 61 | 83 | 62 | 1.17 | 0.98 |
| ***Rn*-index** | 48.9 | 31.9 | 17.5 | 31.2 | 2.79 | 1.02 |
| Composite materials | | | | | | |
| **N (all papers)** | 4011 | 3392 | 11731 | 3271 | 0.34 | 1.04 |
| **top 10%** | 437 | 536 | 1285 | 623 | 0.34 | 0.86 |
| **top 1%** | 75 | 84 | 122 | 87 | 0.61 | 0.97 |
| ***Rn*-index** | 33.3 | 32.9 | 16.5 | 43.1 | 2.02 | 0.76 |

In domestic papers, the results show that only the *Rn*-index reveals the dominance of the USA when attention is focused on the most cited papers (Figures 3 and 4). The top 10% and 1% indicators reflect the much higher number of publications from China rather than the greater impact of the most important papers from the USA. In international collaborative papers, the differences are smaller. This could be expected because, in this type of papers, the apparently independent evaluations for the USA and China are not truly independent. In fact, many of these papers are collaborations between the USA and China and are therefore counted for both countries.

### 4.4. Use of the *Rn*-index instead of percentile indicators

The central idea behind the development of the *Rk*-index was to evaluate the contribution of advanced countries to pushing the boundaries of knowledge and to find a method that does not fail when evaluating Japan (Section 1). In fact, failures of percentile indicators in the evaluation of research in Japan could also occur in other countries. For this reason, the formula for the *Rk*-index was obtained using data for very narrow top percentiles as a reference (Rodríguez-Navarro & Brito, 2024b). However, the same study found that, even at low values of the *Rk*-index, there was always an equivalence with evaluations based on wider top percentiles than those associated with large *Rk*-index values. Specifically, at high values, the *Rk*-index is equivalent to $P_{top\ 0.1\%}$, whereas at low values it is equivalent to $P_{top\ 3\%}$.

For example, in Figure 4, the comparison between the UK and India involves medium and low *Rn* values—0.8 and 8.2, respectively (Table 3). Because in this case the double-rank distribution does not deviate greatly from a power law, it is possible to calculate the top percentile at which the value for the UK is tenfold that for India. As expected from synthetic series, the tenfold difference occurs at the top 0.25%.

These findings indicate that the *Rn*-index could be used as a universal indicator, replacing percentile indicators, and prompted the study of specific cases.

An obvious case is the evaluation of countries that differ greatly in size and efficiency, because scientific success in terms of the number of groundbreaking papers depends on both efficiency and size. A clear example is the comparison between research systems in Singapore and India (Rodríguez-Navarro, 2025a). Table 6 presents, for these two countries, the same data shown for the USA and China in Table 5, along with the ratios between the indicators for Singapore and India.

Table 6. Comparison of the *Rn*-index with the numbers of top 10% and top 1% highly cited papers from Singapore and India for domestic (D) and international collaborative (IC) papers. The last two columns show the ratios between the compared values

| Indicator | Singapore (D) | Singapore (IC) | India (D) | India( IC) | Singapore (D)/ India (D) | Singapore (IC)/ India (IC) |
|---|---|---|---|---|---|---|
| | | | Graphene | | | |
| **N (all papers)** | 686 | 1525 | 3798 | 1688 | 0.18 | 0.90 |
| **top 10%** | 124 | 341 | 184 | 142 | 0.67 | 2.40 |
| **top 1%** | 15 | 46 | 3 | 6 | 5.00 | 7.67 |
| **Rn-index** | 4.2 | 4.5 | 0.4 | 0.7 | 10.5 | 6.43 |
| | | | Semiconductors | | | |
| **N (all papers)** | 275 | 775 | 3059 | 912 | 0.09 | 0.85 |
| **top 10%** | 42 | 177 | 143 | 82 | 0.29 | 2.16 |
| **top 1%** | 6 | 20 | 8 | 6 | 0.75 | 3.33 |
| **Rn-index** | 0.9 | 3.9 | 1.0 | 1.0 | 0.90 | 3.90 |
| | | | Solar cells | | | |
| **N (all papers)** | 365 | 758 | 3141 | 1308 | 0.12 | 0.58 |
| **top 10%** | 48 | 150 | 167 | 86 | 0.29 | 1.74 |
| **top 1%** | 8 | 13 | 3 | 3 | 2.67 | 4.33 |
| **Rn-index** | 1.0 | 2.6 | 0.8 | 0.9 | 1.25 | 2.89 |
| | | | Lithium batteries | | | |
| **N (all papers)** | 317 | 626 | 661 | 381 | 0.48 | 1.64 |
| **top 10%** | 56 | 110 | 19 | 15 | 2.95 | 7.33 |
| **top 1%** | 5 | 15 | 0 | 1 | - | 15.0 |
| **Rn-index** | 1.5 | 5.0 | 0.4 | 0.5 | 3.75 | 10.0 |
| | | | Composite materials | | | |
| **N (all papers)** | 194 | 344 | 2693 | 650 | 0.07 | 0.53 |
| **top 10%** | 35 | 71 | 221 | 85 | 0.16 | 0.84 |
| **top 1%** | 3 | 10 | 11 | 10 | 0.27 | 1.00 |
| **Rn-index** | 0.9 | 5.5 | 1.8 | 3.1 | 0.50 | 1.77 |

The first observation that can be drawn from Table 6 is that $P_{top\ 1\%}$ is not a usefull general indicator for the research assessment in many technological fields because the number of papers at this narrow percentile is too low.

The second observation is that the discrepancy between the *Rn*-index and $P_{top\ 10\%}$ is very large in some cases (Table 6). For example, in the assessment of domestic papers in solar cells, while the *Rn*-indices for the two countries are very similar—0.9 and 1.0—the values of $P_{top\ 10\%}$ are much higher for India than for Singapore—167 versus 48.

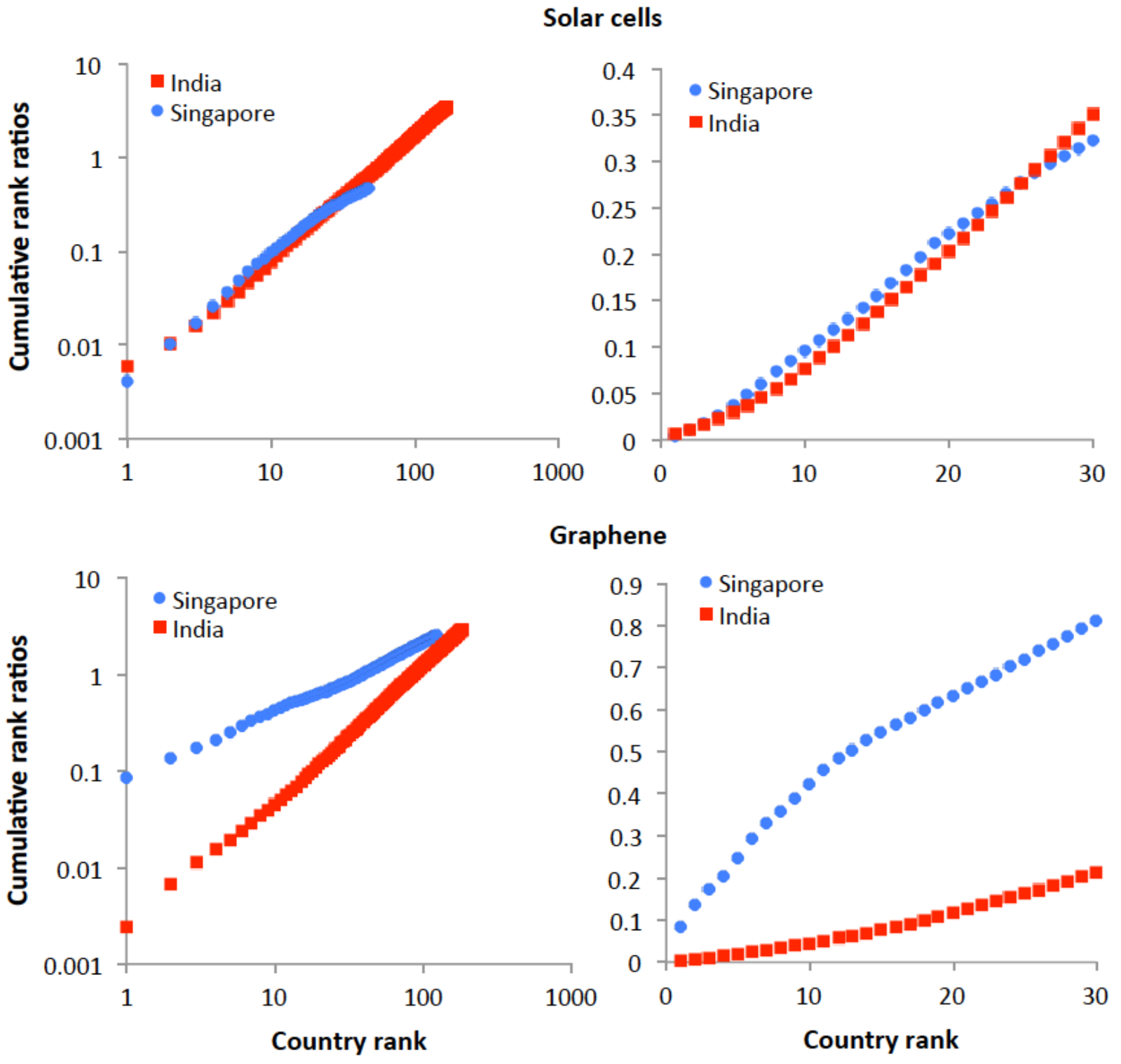

Figure 6. Cumulative values of rank ratios versus country rank of papers from Singapore and India in solar cells and graphene. Left panels: log-log plots of all papers within the top 10% by citations. Right panels: the 30 most cited papers.

For a clearer understanding of these discrepancies, Figure 6 shows the cumulative rank ratios for these two countries in solar cells and graphene—log-log plots of the ranks of papers within the top 10% highly cited papers in left panels and the 30 most cited papers in right panels. In solar cells, the cumulative rank ratios in the two countries almost overlap, indicating similar global ranks of the 30 most cited papers from the two countries. This should be interpreted as evidence of similarity in the highest-quality science produced in the two countries. In contrast, the plot corresponding to the top 10% most cited papers (Figure 6, left panel) indicates that India dominate in the number of papers—167 versus 48—which is irrelevant in terms of the best science.

In the case of graphene domestic papers, while the *Rn*-index is much higher for Singapore than for India—4.2 versus 0.4—the values of $P_{top\ 10\%}$ show the opposite trend—124 versus 184. In Figure 6, the right panel for graphene indicates that, in terms of the highest-quality science in each country, Singapore is clearly ahead of India. The left panel of Figure 6 shows that, as the number of papers considered increases, the differences between Singapore and India decrease. At the top 10% level—$P_{top\ 10\%}$ values of 184 for India versus 124 for Singapore—the values of cumulative rank ratios for all these papers are 2.9 for India and 2.5 for Singapore.

Another interesting case is the evaluation of small countries with reasonable research systems, in which $P_{top\ 1\%}$ is zero in most cases and even $P_{top\ 10\%}$ may be low. The use of the *Rn*-index in these cases would be of great help. For this purpose, I selected Greece and Finland in the same fields as Singapore and India (Table 6), excluding lithium batteries because the numbers of papers from these countries are too low. Table 7 summarizes the data, and Figure 7 shows the cumulative plots of rank ratios. The data in Figure 7 indicate that both countries are very similar in semiconductors and solar cells, a conclusion that can be reached using either $P_{top\ 10\%}$ or the *Rn*-index (Table 7). In the case of composite materials, the dominance of Greece is evident (Figure 7) and is probably revealed more accurately by the *Rn*-index than by $P_{top\ 10\%}$.

Table 7. Comparison of the value of the *Rn*-index with the number of domestic top 10% highly cited papers from Greece and Finland in four technological fields

| | **Graphene** | | **Semiconductors** | | **Solar cells** | | **Composite materials** | |
|---|---|---|---|---|---|---|---|---|
| | **Greece** | **Finland** | **Greece** | **Finland** | **Greece** | **Finland** | **Greece** | **Finland** |
| **N (all papers)** | 184 | 95 | 147 | 112 | 216 | 128 | 195 | 111 |
| **top 10%** | 8 | 6 | 12 | 6 | 17 | 10 | 15 | 9 |
| ***Rn*-index** | 0.11 | 0.10 | 0.14 | 0.11 | 0.15 | 0.13 | 0.42 | 0.17 |

The case of graphene in Greece has certain similarity with the case of Japan in semiconductors (Figure 2); the rank of the first paper is much smaller than those of the following papers—266 versus 3,862 and 4,829—which does not occur in Finland. Figure 7, in addition to showing the actual cumulative plots of rank ratios, also includes a plot in which the actual rank of the first paper has been replaced by a value predicted from the remaining ranks. In this simulation, the *Rn*-index for Greece decreases from 0.11 to 0.08.

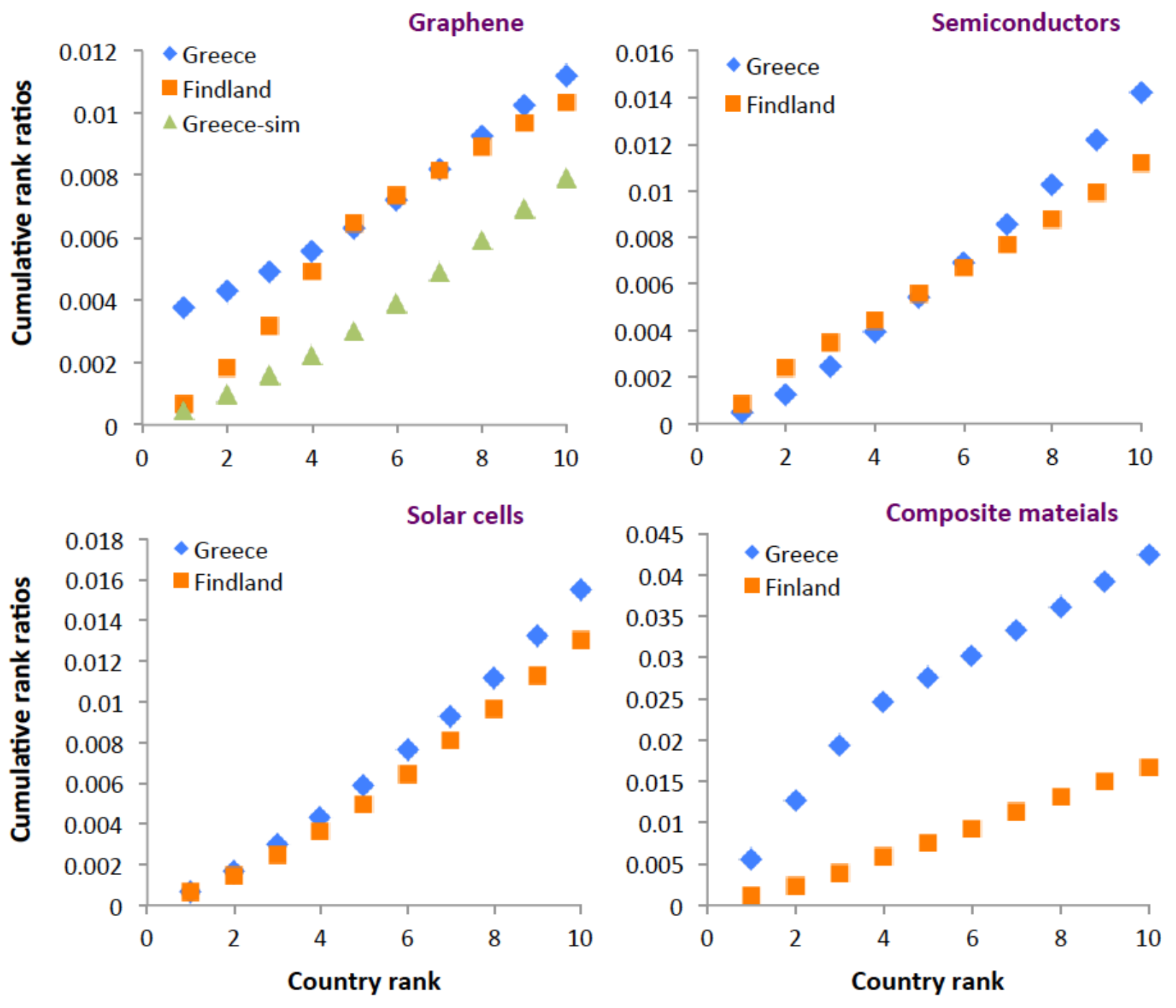

Figure 7. Cumulative values of rank ratios versus country rank of the 10 most cited papers from Greece and Finland in four technological fields. The graphene panel includes a simulation in which the rank value of the first paper from Greece, 266, was changed to 2100.

## 5. Discussion

### 5.1. Aim of the *Rn*-index

Sections 1.1 and 1.2 highlight the challenge of developing research indicators specifically suited to evaluating landmark research that pushes the boundaries of knowledge. A similar challenge arises when evaluating the best science produced by countries. Figure 4 shows that there are at least two ways in which one country can contribute more to the progress of science than another country that publishes a larger number of papers: through higher efficiency with low deviations from an ideal power-law model, or through significant deviations from the ideal model in the most cited papers. In international comparisons of countries conducting groundbreaking research, the challenge is to evaluate the best publications of each country—those that contribute most to the advancement of knowledge. Internally, each country should investigate its own performance by considering the ranks of all papers. For example, why is South Korea ahead of the EU and the UK ahead of India in terms of groundbreaking papers (Figure 4)?

The *Rn*-index—along with its predecessor, the *Rk*-index (Rodríguez-Navarro & Brito, 2024b)—addresses this challenge by using the global ranks of the most cited papers. Similarly to top percentile indicators, this nonparametric method implies a normalization regarding citation differences across fields.

Unlike the *Rk*-index, whose formula is complex—having been obtained through comparison with very narrow top-percentile indicators in synthetic series—the *Rn*-index has a very simple formula: the sum of rank ratios, obtained by dividing the country rank by the global rank (Table 1). At the high values of the *Rk*-index that apply to the USA and China, the *Rn*-index is not affected by the low resolution of its predecessor, including its noncompliance with the combination-addition property of research indicators (Tables 1 and 2).

The properties of the cumulative values of rank ratios raise the question of the most appropriate number of papers to consider when calculating the *Rn*-index. Based on the scatter plots in Figure 3, using more than 10 papers can be ruled out because the purpose of the index is to evaluate the best science of each country, and considering many papers decreases sensitivity regarding the best science. For this purpose, the five-paper variant of the *Rn*-index may be more effective than the 10-paper version (Table 3). However, in most cases the differences are not important, and using 10 papers may increase the stability of the indicator.

The *Rn*-index is very sensitive to extreme cases—such as Japan in semiconductors (Figure 2, Table 4) and Greece in graphene (Figure 7)—in which the rank of the most cited paper is much lower than expected from the ranks of the following papers. This raises the question of how much weight should be assigned to these papers. Using the geometric mean of rank ratios instead of the *Rn*-index, which is equivalent to the arithmetic mean, greatly reduces the impact of such outliers. However, it is still unclear whether such a correction is desirable because it also affects countries with less important deviations (Table 4), and it is not clear whether the weight assigned to the most cited papers should be reduced. For example, in the case of South Korea's domestic papers in solar cells (Figures 1, 3, and 4), the difference between the geometric mean and the *Rn*-index is not negligible—22.1 versus 17.1 (Table 4).

These cases raise the question of the scientific significance of outlier papers that, mathematically, cannot be considered part of the tail defined by the ranks of the subsequent papers. In the case of the most cited paper globally, as with Japan in semiconductors, one interesting observation is that such extreme cases tend to occur in very advanced countries: the USA, China, Japan, South Korea, and a few others. Apparently, these cases do not occur in countries with less competitive research systems, raising a deeper philosophical issue: the role of "accident" versus "sagacity" in serendipitous scientific discoveries (e.g., Campanario, 1996; Ross et al., 2024). One technical uncertainty in these cases is that the exceptional paper might be a review paper that has not been classified as such by the database and has therefore been included among the retrieved papers. This risk is evident and should be monitored manually. In a previous paper (Rodríguez-Navarro, 2025b), statistical papers and treatment guidelines in the field of cancer were eliminated manually.

More frequent may be cases such as Greece in graphene (Figure 7), in which the outlier paper is far from the top ranks, although the considerations remain the same.

Possibly the best solution for all these cases, and as a general rule, is to calculate the *Rn*-index independently for each year in a series of years—e.g., four—and calculate the mean.

### 5.2. Considerations on the *Rn*-index

A distinguishing feature of the *Rn*-index is its conceptual and computational simplicity: it is simply the sum of the ratios between the country and global ranks of its 10 best-ranked papers by citation (Table 1). This simplicity is advantageous for research indicators, as their societal impact on research policy depends on policymakers, whose disconnection from academia in many cases (Jolivet, 2024) may be reduced if evaluation approaches are straightforward.

Like $P_{top\ 1\%}$ and $P_{top\ 10\%}$, the *Rn*-index is a size dependent indicator, and transforming it into a size-independent indicator by dividing it by the total number of publications would result in an uncertain indicator (Rodríguez-Navarro, 2025b). Alternatively, in order to estimate the research efficiency of a country, the *Rn*-index may be normalized by the GNP, considering GDP per capita where appropriate.

The evaluation of international collaborative papers presents a challenge for all research indicators (Olechnicka et al., 2019; Zanotto et al., 2016), and this challenge is even greater assessing disruptive research. Although fractional counting—as proposed for the *h*- and *g*-indices (Egghe, 2008)—could theoretically be applied to the *Rn*-index, its application is not advisable. Fractional counting does not fully correct the inflated evaluation of less competitive countries and undermines the measured performance of dominant ones (Zanotto et al., 2016). Therefore, while it is essential to complement the evaluation of domestic papers with information on international collaborations, such information cannot be reliably obtained through either full or fractional counting. Instead, it must be assessed on a case-by-case basis using specific methods that distinguish the collaborations with different countries (Rodríguez-Navarro, 2025a; Zanotto et al., 2016).

The *Rn*-index values calculated in this study correspond to individual technologies as is done by the Australian Strategic Policy Institute (Gaida et al., 2023; Leung et al., 2024). The applicability the *Rn*-index to broader fields—such as engineering, which encompasses multiple technologies—is uncertain. It remains unclear whether such an application is both necessary and feasible. Further research and discussion would be required should this approach be pursued.

### 5.3. The *Rn*-index versus top percentiles

From a conceptual point of view, the *Rn*-index and percentile indicators are completely different. Considering the list of global papers ordered from highest to lowest number of citations, the latter correspond to the number of country papers that rank below certain global rank thresholds. In contrast, the former is calculated from the global and country ranks of a fix number of country papers.

Despite these conceptual differences, a previous study using the *Rk*-index with synthetic series (Rodríguez-Navarro & Brito, 2024b), together with the findings of this study (Section 4.3), indicate that the *Rn*-index can be used instead of top percentiles as a general indicator of the best science produced by a country.

Currently, the research evaluation of countries is usually performed using the number of top 10% or top 1% highly cited papers (e.g., King, 2004; Gaida et al., 2023; Leung et al., 2024; National Institute of Science and Technology, 2022; Wagner et al., 2022). However, these indicators do not reflect the capacity or performance of countries at the level of landmark papers and are very sensitive to differences in country sizes—in terms of publications.

When comparing two countries that are not very different in size, have similar efficiency, and produce only "normal" science, $P_{top\ 1\%}$ and $P_{top\ 10\%}$ may be reasonable indicators, but in other cases, the comparison may be highly misleading. Tables 5 and Figure 4 demonstrate that $P_{top\ 1\%}$ and $P_{top\ 10\%}$ are excessively influenced by size and are insensitive to the production of elite groups of researchers that may have a high probability of making important advances.

This study shows that the *Rn*-index provides a reasonable solution for evaluating the best science produced by each country: both those producing most of the important discoveries—the USA and China—and those with a lower or much lower probability of contributing important discoveries. However, because this study has focused on countries, the question arises as to whether the *Rn*-index may also be more suitable than $P_{top\ 1\%}$ and $P_{top\ 10\%}$ for the evaluation of institutions, at least for the most productive institutions, such as those selected in the *Leiden Ranking* (https://www.leidenranking.com/).

Although a study focused on institutions is still pending, the simplicity of the *Rn*-index calculation, its support from both real cases and synthetic series simulating citations, and the inconsistencies of percentile indicators strongly suggest that this index may be the best option.

## 6. Concluding remarks

Assessing and comparing the research outputs of different countries is a complex task. In some cases, citation-based patterns can be described mathematically, whereas in others they exhibit discontinuities that may reflect the heterogeneity of the research community. Under such conditions, a single top-percentile indicator is insufficient to capture differences in research performance across countries, and different percentile indicators may even lead to conflicting conclusions (Figure 4). The *Rn*-index is designed to evaluate the highest-quality scientific contributions produced by a country, whether they represent groundbreaking discoveries or high-level mainstream research. Nevertheless, the relationship between these outstanding contributions and the broader body of scientific output must be examined individually for each country.

## CONFLICT OF INTEREST

The author declares that there is no conflict of interest.